# Spin-phonon coupling in Gd(Co$_{1/2}$Mn$_{1/2}$)O$_3$ perovskite

R. X. Silva[1,2], H. Reichlova[3,4], X. Marti[3,4,5], D. A. B. Barbosa[1,7], M. W. Lufaso[7], A. P. Ayala[8] and

C. W. A. Paschoal[2,5,6*]

[1] Curso Ciências Naturais, Universidade Federal do Maranhão, Campos VII, 65400-000, Codó-MA, Brazil.

[2] Departamento de Física, Universidade Federal do Maranhão, Campus do Bacanga, 65085-580, São Luis-MA, Brazil.

[3] Institute of Physics ASCR, v.v.i., Cukrovarnická 10, 162 53 Praha 6, Czech Republic

[4] Faculty of Mathematics and Physics, Charles University, Ke Karlovu 5, 12116 Praha 2, Czech Republic

[5] Department of Materials Science and Engineering, University of California Berkeley, Berkeley - CA, 94720-7300, United States.

[6] Department of Physics, University of California Berkeley, 94720-7300, Berkeley - CA, United States

[7] Department of Chemistry, University of North Florida, 1 UNF Drive, Jacksonville, FL 32224, United States.

[8] Departamento de Física, Universidade Federal do Ceará, Campus do Pici, PO Box 6030, 60455-970, Fortaleza - CE, Brazil.

## Abstract

We have investigated the temperature-dependent Raman-active phonons and the magnetic properties of Gd(Co$_{1/2}$Mn$_{1/2}$)O$_3$ perovskite ceramics in the temperature range from 40 K to 300 K. The samples crystallized in an orthorhombic distorted simple perovskite, whose symmetry belongs to the *Pnma* space group. The data reveals spin-phonon coupling near the ferromagnetic transition occurring

---

[*] Corresponding Author: C. W. A Paschoal: e-mail: paschoal@ufma.br; Tel +55 98 3301 8291

at around 120 K. The correlation of the Raman and magnetization data suggests that the structural order influences the magnitude of the spin-phonon coupling.

Rare-earth manganite perovskites present a very rich interplay between electric and magnetic properties that has been demonstrated in a plethora of phenomena observed in either bulk, powders or strained thin films. For instance, we can cite giant and colossal magnetoresistances observed in $LaMnO_3$-based perovskites [1], occurrence of multiferroic behavior [2,3] and the discovery of the strong magnetodielectric coupling in ordered $La(Ni_{1/2}Mn_{1/2})O_3$ perovskite (usually written as $La_2NiMnO_6$, hereafter LNMO) [4]. The exploration of the chemical substitutions in the B-site has concentrated many efforts in $REMnO_3$ (RE = rare earth) perovskites, since the choice of magnetic and ionic radius of the RE and substitute B cations, drives the magnetism, the ferroelectricity, and their coupling [5–12].

The ferroic and multiferroic properties depend on the unit cell topology (cations' size, angles and distances), therefore investigations of the spin-phonon coupling of manganites are of crucial interest and a convenient tool to study the magnetically induced ferroelectric systems [13]. Indeed, there is a vast literature on the investigation of spin-phonon coupling in magnetoelectric materials using both Raman and infrared spectroscopies comprising both simple and double perovskites [7,12,14–16]. Sample preparation, including the exploitation of epitaxy-induced strains, impacts the multiferroic properties and raises interesting differences between specimens prepared following different routes but with identical chemical composition. A paradigmatic example is the Co-substituted ordered manganite $La_2CoMnO_6$ (LCMO) that shows spin-phonon coupling in thin-films, while the phonon modes remain insensitive to magnetic order in bulk samples [17]. The conclusive explanation of this particular difference remains unraveled and calls for further investigations to the entire family, including ordered (conventionally typed $RE_2CoMnO_6$) and disordered (conventionally typed $RE(Co_{1/2}Mn_{1/2}O_3)$) compounds.. In this Letter, we have investigated the spin-phonon coupling in the Gd(Co$_{1/2}$Mn$_{1/2}$)O$_3$ (GCMO) perovskite using temperature-dependent Raman spectroscopy, which has been an important tool to investigate order, magnetic transitions, and spin-phonon coupling in perovskites [12,15,18–22].

Polycrystalline samples of GCMO were synthesized by polymeric precursor method [23] using cobalt acetate tetrahydrate (C$_4$H$_6$CoO$_4$·4H$_2$O, Sigma Aldrich), manganese nitrate hydrate (MnN$_2$O$_6$·xH$_2$O, Sigma Aldrich) and gadolinium oxide (Gd$_2$O$_3$, Sigma, Aldrich) as high purity metals sources. The resin and puff (powder obtained by the resin pre-calcination) were obtained following the steps described elsewhere [21]. The puff was lightly ground using an agate mortar and calcined at 1100 °C for 16h to obtain GCMO sample. The crystalline structure of the sample was probed by powder X-ray diffraction using a Rigaku Ultima III with Cu-K$\alpha$ radiation (40 kV, 34 mA) over a range from 10° to 115° (0.03°/step with 10s/step). The power XRD pattern was compared with data from ICSD (Inorganic Crystal Structure Database, FIZ Kalsruhe and NIST) International diffraction database (ICSD# 23562) [24]. The structure was refined using the Fullprof Suite v.2.05 [25]. Raman spectroscopy measurements were performed using a Jobin-Yvon T64000 Triple Spectrometer configured in a backscattering geometry coupled to an Olympus Microscope model BX41 with a 20x achromatic lens. The 514.5 nm line of an Innova Coherent laser operating at 100 mW was used to excite the signal, which was collected in a N$_2$-cooled CCD detector. All slits were set up to achieve a resolution lower than 1 cm$^{-1}$. Low-temperature measurements were performed by using a closed-cycle He cryostat where the temperature was controlled to within 0.1 K. Magnetic measurements were carried out in Quantum Design (QD) superconducting quantum interference device (SQUID). Temperature sweeps were collected with 4 cm long Reciprocating Sample Option scans.

In Figure 1 we show the powder X-ray pattern obtained for the GCMO sample. The reflections were indexed according to a disordered orthorhombic perovskite lattice, belonging to the *Pnma* space group. The refined structure parameters are in excellent agreement with those obtained previously by solid state reaction [26] and theoretically calculated by the SPuDS program [27]. The refinement and structural parameters are summarized in Tables S1, S2, S3 and S4 of the supplementary material. Virtually unavoidable traces of $Gd_2O_3$ (<0.5%) were detected as secondary phase. $RE_2O_3$ is routinely observed in the synthesis of rare-earth mixed perovskites [12,14].

Figure 2(a) shows the room-temperature Raman spectrum corresponding to GCMO sample. According to the ionic site occupation, group theory predicts 24 Raman-active phonons in GCMO, whose distribution into the irreducible representation of the $mmm$ factor group is $7A_g \oplus 5B_{1g} \oplus 7B_{2g} \oplus 5B_{3g}$ (see Table S5 of the supplementary material) [30]. We observed 10 modes at room temperature. The number of observed modes, lower than the predictions, is not surprising since $GdMnO_3$ and $GdCoO_3$ show low-intensity Raman spectra, even at low temperatures [13,29]. The observed phonons were classified based on GF Wilson matrix method [30–32]. The modes can be divided into a lower (below 450 cm$^{-1}$) and higher (above 450 cm$^{-1}$) wavenumber regions, separated at 450 cm$^{-1}$. In the low wavenumber region, the vibrations are usually not pure, comprising combinations of pure vibrations (stretching, bending, tilt, etc.) with large displacements of the Gd ion. In the high wavenumber region, the vibrations are almost pure. The stronger phonon at approximately 628 cm$^{-1}$ corresponds to the (Mn,Co)O$_6$ octahedra symmetrical stretching, but it is mainly due to the equatorial oxygen ions (O2 ions). This is in excellent agreement with preceding works by Iliev et al [33]. The mode at 461 cm$^{-1}$ is a (Mn,Co)O$_6$ octahedra bending. In addition, the (Mn,Co)O$_6$ octahedra tilt along the y axes and other bending mode were observed at 293 and 494 cm$^{-1}$, respectively. These modes are depicted in Figure 2(a). In addition to these modes, other weak phonons were observed at 251, 278, 330, 391, 440, and 531 cm$^{-1}$. A full classification of the phonons observed in GCMO is listed in Table 1. The classification agrees very well with those proposed for other Mn-based perovskites with $Pnma$ symmetry [13,33] and with previous GCMO results [34]. A detailed description of the calculations used to model the phonons and pictures of all the observed Raman-active phonons are given in the supplementary material.

Figure 2(b) shows the temperature-dependent Raman spectra of GCMO between the room temperature and 40 K. As expected, no remarkable changes were observed since GCMO does not undergo a structural phase transition in this temperature range [26]. The temperature-dependence of the (Mn,Co)O$_6$ stretching and bending wavenumber positions is shown in Figure 3. The stretching shows a deviation from the usual anharmonic contribution to the temperature dependence of the phonon wavenumber, which was modeled by Balkanski et al [35] as:

$$\omega(T) = \omega_o - C\left[1 + \frac{2}{(e^{\hbar\omega_o/k_BT} - 1)}\right]$$

with $C$ and $\omega_o$ being fitting parameters. In the absence of structural phase transitions, as it is found in GCMO, the temperature dependence of the phonons follows this behavior. This deviation is similar to that observed in other perovskites with magnetic transitions, and it is associated to the renormalization of the phonons induced by the magnetic ordering [7,12,15,17,36-38], usually understood as a signature of coupling between the spin and phonons. To check this assumption, we measured, in the same temperature range, the spontaneous magnetization of GCMO, which is shown in Figure 3. The magnetization curve reveals the onset of the net magnetic moment at the Curie temperature near 120 K, which is consistent with other magnetic measurements in GCMO [26]. Our data evidence that the behavior of the stretching mode departs from the anharmonic model near this transition, confirming the coupling between the magnetic ordering and this phonon. The comparison between the departure from the anharmonic behavior of the stretching and the square of the magnetization, plotted in the inset of the Figure 3(a), highlights that the phonon renormalization scales well with the square of the magnetization, as predicted the mean field theory [39].

Our major finding is that, contrary to similar perovskites [7,12,14,15], the renormalization implied in a hardening of the phonon, suggesting that the spin-phonon, in the present case, does not stabilize the transition [39]. This behavior is different of that observed in the stretching mode in $GdMnO_3$ simple perovskite phonons [13] and akin to that observed in $GdCoO_3$ [40]. Ultimately, it reflects a large coupling of the magnetic ordering coupling with the symmetrical stretching in $GdCoO_3$ when compared with $GdMnO_3$ [11,40].

On the other hand, it is interesting to compare the LCMO and GCMO spin-phonon coupling properties. LCMO does not exhibit such a coupling in bulk, but it was reported in thin films [18]. However, this effect is clearly observed in GCMO (Figure 3(a)). The main differences are: the structural order (LCMO and GCMO are ordered and disordered perovskites, respectively); the ionic radii of the rare earth ion (Gd is almost 10% smaller than La); and the rare earth magnetic moment (presented by Gd, but absent in La). It is also important to take into consideration that the simple perovskites $LaMnO_3$ and $LaCoO_3$ exhibit spin-phonon coupling [39,40]. However, It was proved that in double perovskites this effect is not fully driven by the ionic radii size [12]. Based on these results, the observation of spin-phonon coupling in GCMO suggests that the magnitude of the spin-phonon can be turned by the structural order. As a result, we anticipate that GCMO thin films should exhibit large spin-phonon coupling than LCMO, since the effect is rather observed in bulk in GCMO.

In summary, we have investigated the temperature dependence of the Raman spectra of the $Gd(Co_{1/2}Mn_{1/2})O_3$ perovskite between 40 K and 300 K. Magnetic characterization of the same samples indicated a ferromagnetic transition at $T_c$ ~120 K. While no remarkable spectral changes in Raman spectra could be ascribed to structural phase transitions, the temperature dependence of the symmetric stretching of the oxygen octahedra exhibits an anomalous hardening below the ferromagnetic transition. We argue that this effect is related to the phonon renormalization induced by the spin-phonon coupling, suggesting that the structural ordering influences in the magnitude of this phenomenon.

## Acknowledgments

The Brazilian authors acknowledge the partial financial support from CNPq, CAPES, FUNCAP and FAPEMA Brazilian funding agencies. C. W. A. Paschoal acknowledges R. Ramesh for all support at Univ. California Berkeley. D. A. B. Barbosa acknowledges to the CAPES Scholarship Grant n° 10423-12-5 (PDSE 2012). M. W. Lufaso acknowledges UNF for the Munoz Professorship. X. Marti acknowledges the Grant Agency of the Czech Republic No. P204/11/P339.

**References**

[1] A. Urushibara, T. Arima, A. Asamitsu, G. Kido, and Y. Tokura, Physical Review B **51**, 14103 (1995).

[2] I. Fina, L. Fàbrega, X. Martí, F. Sánchez, and J. Fontcuberta, Applied Physics Letters **97**, 232905 (2010).

[3] T. Kimura, T. Goto, H. Shintani, K. Ishizaka, T. Arima, and Y. Tokura, Nature **426**, 55 (2003).

[4] N. S. Rogado, J. Li, A. W. Sleight, and M. A. Subramanian, Advanced Materials **17**, 2225 (2005).

[5] D. Choudhury, P. Mandal, R. Mathieu, A. Hazarika, S. Rajan, A. Sundaresan, U. Waghmare, R. Knut, O. Karis, P. Nordblad, and D. Sarma, Physical Review Letters **108**, (2012).

[6] S. Baidya and T. Saha-Dasgupta, Physical Review B **84**, 035131 (2011).

[7] K. Truong, M. Singh, S. Jandl, and P. Fournier, Physical Review B **80**, 134424 (2009).

[8] S. M. Zhou, Y. Q. Guo, J. Y. Zhao, S. Y. Zhao, and L. Shi, Applied Physics Letters **96**, 262507 (2010).

[9] Y. Bai, X. Liu, Y. Xia, H. Li, X. Deng, L. Han, Q. Liang, X. Wu, Z. Wang, and J. Meng, Applied Physics Letters **100**, 222907 (2012).

[10] S. Kazan, F. A. Mikailzade, M. Özdemir, B. Aktaş, B. Rameev, A. Intepe, and A. Gupta, Applied Physics Letters **97**, 072511 (2010).

[11] S. Lv, H. Li, Z. Wang, L. Han, Y. Liu, X. Liu, and J. Meng, Applied Physics Letters **99**, 202110 (2011).

[12] R. B. Macedo Filho, A. P. Ayala, and C. W. A. Paschoal, Applied Physics Letters **102**, 192902 (2013).

[13] W. Ferreira, J. Agostinho Moreira, A. Almeida, M. Chaves, J. Araújo, J. Oliveira, J. Machado Da Silva, M. Sá, T. Mendonça, P. Simeão Carvalho, J. Kreisel, J. Ribeiro, L. Vieira, P. Tavares, and S. Mendonça, Physical Review B **79**, 054303 (2009).

[14] H. S. Nair, D. Swain, H. N., S. Adiga, C. Narayana, and S. Elzebeth, Journal of Applied Physics **110**, 123919 (2011).

[15] M. N. Iliev, H. Guo, and A. Gupta, Applied Physics Letters **90**, 151914 (2007).

[16] X. Luo, Y. P. Sun, B. Wang, X. B. Zhu, W. H. Song, Z. R. Yang, and J. M. Dai, Solid State Communications **149**, 810 (2009).

[17] M. Iliev, M. Abrashev, A. Litvinchuk, V. Hadjiev, H. Guo, and A. Gupta, Physical Review B **75**, (2007).

[18] K. Truong, J. Laverdière, M. Singh, S. Jandl, and P. Fournier, Physical Review B **76**, 132413 (2007).

[19] S. Zhao, L. Shi, S. Zhou, J. Zhao, H. Yang, and Y. Guo, Journal of Applied Physics **106**, 123901 (2009).

[20] Z. Zhang, H. Jian, X. Tang, J. Yang, X. Zhu, and Y. Sun, Dalton Transactions (Cambridge, England : 2003) **41**, 11836 (2012).

[21] J. E. F. S. Rodrigues, E. Moreira, D. M. Bezerra, A. P. Maciel, and C. W. de Araujo Paschoal, Materials Research Bulletin **48**, 3298 (2013).

[22] A. Dias, L. A. Khalam, M. T. Sebastian, C. W. A. Paschoal, and R. L. Moreira, Chemistry of Materials **18**, 214 (2006).

[23] M. P. Pechini, U.S. Patent No. 3.330.697 (1967).

[24] A. Marsh and C. C. Clark, Philosophical Magazine **19**, 449 (1969).

[25] J. Rodríguez-Carvajal, Physica B: Condensed Matter **192**, 55 (1993).

[26] X. L. Wang, J. Horvat, H. K. Liu, A. H. Li, and S. X. Dou, Solid State Communications **118**, 27 (2001).

[27] M. W. Lufaso and P. M. Woodward, Acta Crystallographica. Section B, Structural Science **57**, 725 (2001).

[28] D. L. Rousseau, R. P. Bauman, and S. P. S. Porto, Journal of Raman Spectroscopy **10**, 253 (1981).

[29] W. Wei-Ran, X. Da-Peng, S. Wen-Hui, D. Zhan-Hui, X. Yan-Feng, and S. Geng-Xin, Chinese Physics Letters **22**, 2400 (2005).

[30] L. Piseri and G. Zerbi, Journal of Molecular Spectroscopy **261**, 254 (1968).

[31] T. Shimanouchi, M. Tsuboi, and T. Miyazawa, The Journal of Chemical Physics **35**, 1597 (1961).

[32] E. Dowty, Physics and Chemistry of Minerals **14**, 67 (1987).

[33] M. N. Iliev, H.-G. Lee, V. N. Popov, Y. Y. Sun, C. Thomsen, R. L. Meng, and C. W. Chu, Physical Review B **57**, 2872 (1998).

[34] C. L. Bull and P. F. McMillan, Journal of Solid State Chemistry **177**, 2323 (2004).

[35] M. Balkanski, R. Wallis, and E. Haro, Physical Review B **28**, 1928 (1983).

[36] M. Iliev, P. Padhan, and A. Gupta, Physical Review B **77**, 172303 (2008).

[37] M. N. Iliev, M. M. Gospodinov, M. P. Singh, J. Meen, K. D. Truong, P. Fournier, and S. Jandl, Journal of Applied Physics **106**, 023515 (2009).

[38] K. D. Truong, M. P. Singh, S. Jandl, and P. Fournier, Journal of Physics. Condensed Matter : an Institute of Physics Journal **23**, 052202 (2011).

[39] E. Granado, A. García, J. Sanjurjo, C. Rettori, I. Torriani, F. Prado, R. Sánchez, A. Caneiro, and S. Oseroff, Physical Review B **60**, 11879 (1999).

[40] A. Ishikawa, J. Nohara, and S. Sugai, Physical Review Letters **93**, 136401 (2004).

# Tables

Table I – Assignment of the Gd(Co$_{1/2}$Mn$_{1/2}$)O$_3$ Raman-active modes based on the GF Wilson matrix analysis.

| Observed (cm$^{-1}$) | Calculated (cm$^{-1}$) | Symmetry | Main atomic motions |
|---|---|---|---|
| 251 | 255.6 | A$_g$ | Octahedra Bending / RE moves in *xz* plane |
| 278 | 277.9 | B$_{1g}$ | octahedra out-of-phase tilt along *y* axis / RE and apical O move along *y* axis |
| 293 | 295.9 | A$_g$ | octahedra in-phase tilt along *y* axis / RE and apical O move in *xz* plane |
| 330 | 334.5 | B$_{1g}$ | octahedra antisymmetric stretching / RE moves along *y* axis |
| 391 | 381.1 | A$_g$ | octahedra antisymmetric stretching / RE moves in *xz* plane |
| 440 | 439.5 | B$_{2g}$ | octahedra out-of-phase tilt along *x* axis / RE and apical O move in *xz* plane |
| 461 | 463.1 | A$_g$ | Octahedra bending |
| 494 | 494.2 | A$_g$ | Octahedra bending |
| 531 | 540.4 | B$_{2g}$ | Octahedra antisymmetric stretching |
| 628 | 624.8 | B$_{2g}$ | Octahedra symmetric stretching in plane *xz* due to the equatorial oxygens / RE moves in the same plane |

# Figures

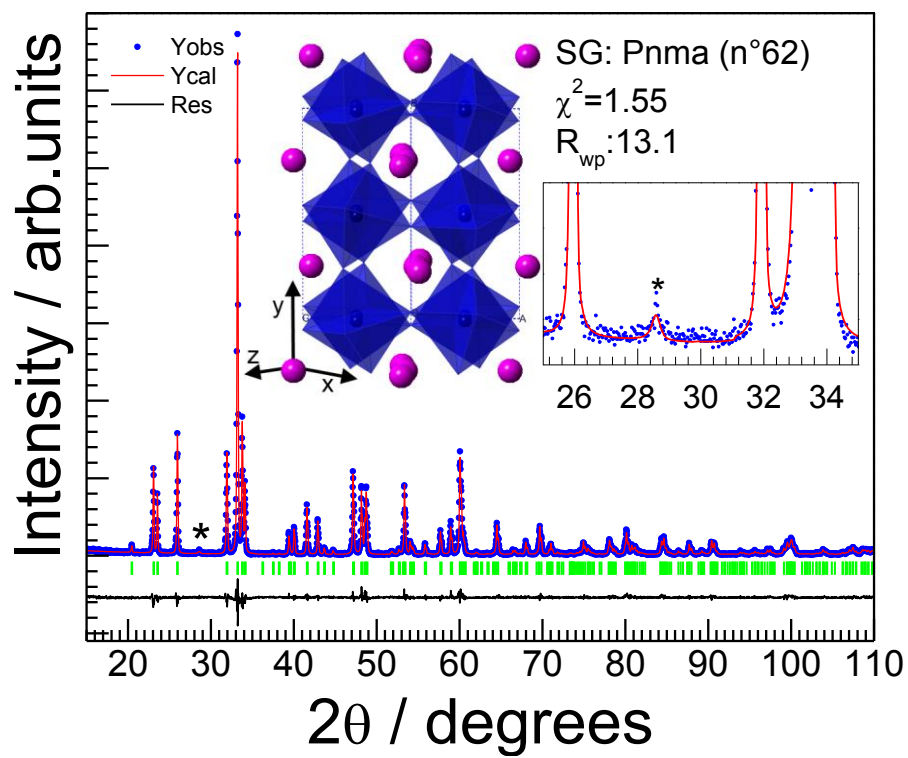

**Figure 1**–Powder X-ray diffraction pattern of GCMO ceramic sintered at 1100 °C/16 h. The red solid line is the fitting using the Rietveld method and the black line is the residual between the experimental and calculated patterns. The inset on right shows a detailed view of the peaks. On the upper left a view of the structure is shown. The asterisk shows the most intense peak associated with the $Gd_2O_3$ secondary phase.

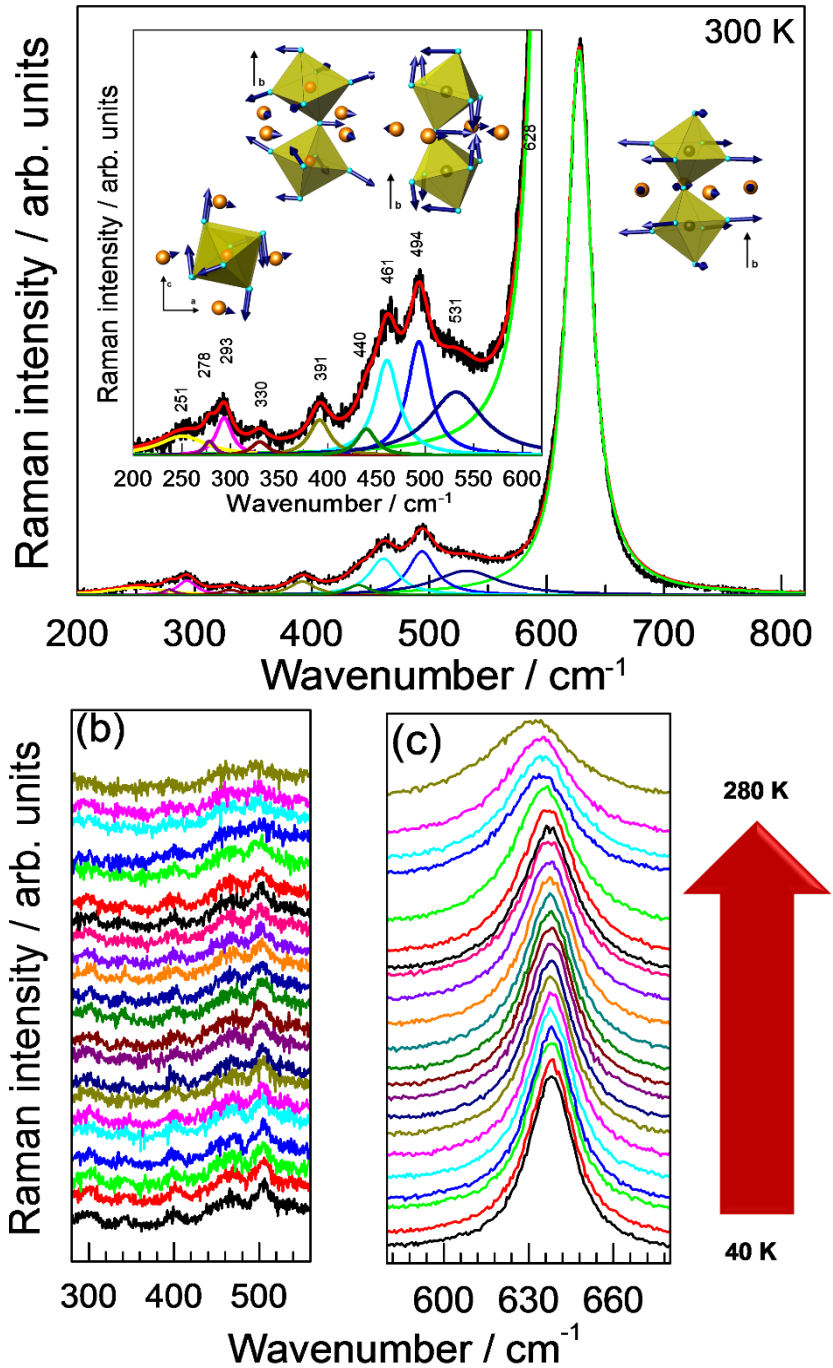

**Figure 2** – Top: Raman spectrum of GCMO at room temperature. The inset shows the low-wavenumber region of the spectrum. The atomic displacements of the main phonons observed are depicted: 626 cm$^{-1}$, 494 cm$^{-1}$, 391 cm$^{-1}$ and 293 cm$^{-1}$; Bottom: temperature-dependent Raman-active phonon spectra of GCMO in the wavenumber range of (b) low-wavenumber region and (c) high-wavenumber (stretching region). The temperature

measurements were performed in temperature steps of 20 K far from the magnetic transition and of 5 K near the transition.

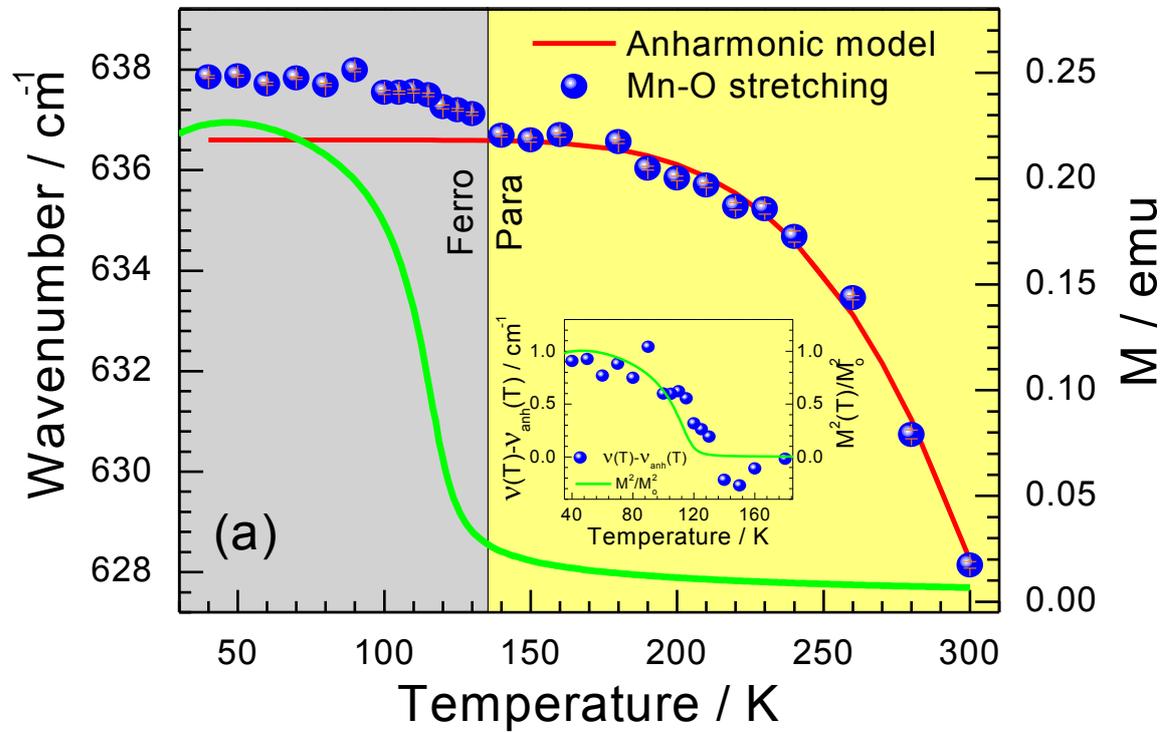

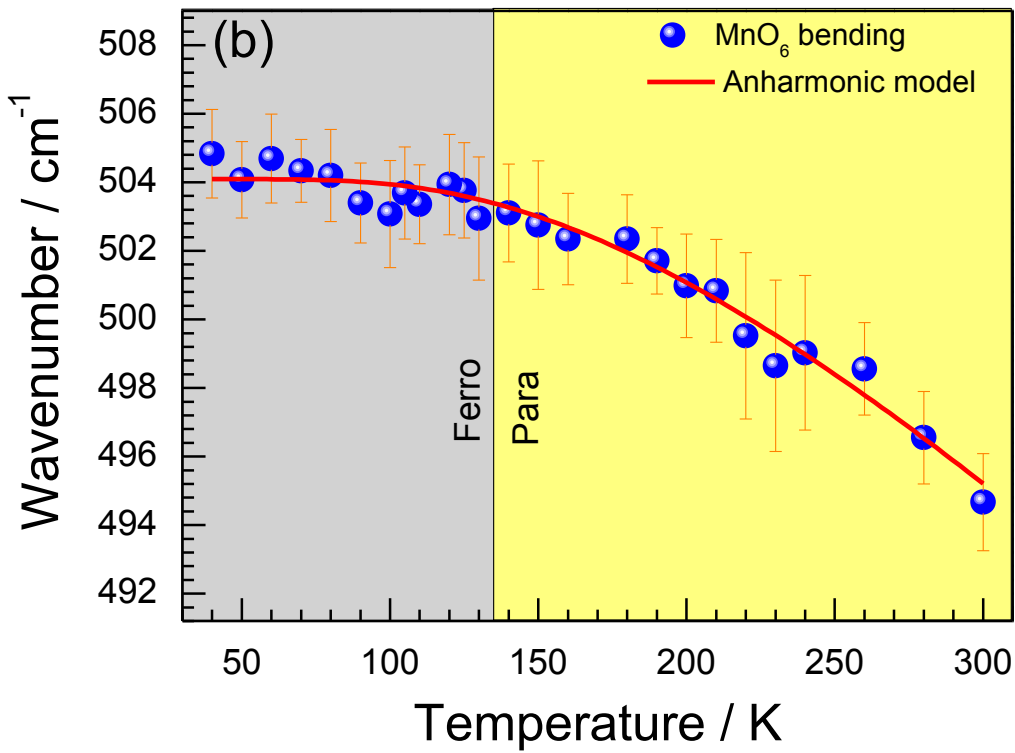

**Figure 3** - Temperature dependence of the phonon positions of two typical modes observed for GCMO (blue spheres). The red lines show the anharmonic effect contribution to the temperature dependence of the phonons modeled by the Balkanski model. Figure (a) shows the magnetization (green line) obtained for GCMO. The inset in figure (a) shows the temperature dependence of the departure from the anharmonic behavior of the mode that exhibits the spin-phonon coupling compared with $(M(T)/M_o)^2$.